\documentclass[multphys,vecphys]{svmult}

\usepackage{graphicx}    
\usepackage{multicol}    

\begin{document}

\title*{Cosmic Acceleration, Scalar Fields and Observations}
\author{C\'esar A. Terrero-Escalante}
\institute{Instituto~de~F\'{\i}sica,~UNAM,
~Apdo.~Postal~20-364,~01000,~M\'exico~D.F.,~M\'exico.
\texttt{cterrero@fis.cinvestav.mx}}

%
%
\maketitle

\begin{abstract}
Conditions for
accelerated expansion of
Friedmann--Robertson--Walker space--time are analyzed. Connection of this
scenario with present--day observations are reviewed.
It is explained
how a scalar field could be responsible for cosmic
acceleration observed in present times and predicted for the very early
Universe. Ideas
aimed at answering whether is that the actual case for our
Universe are described.
\end{abstract}

\section{Introduction}
\label{sec:intro}

The Standard Big Bang Model (SBB),
based on a Friedmann--Robertson--Walker (FRW)
Universe, evolves in time with essentially two phases. In the first one
the energy related to relativistic matter (known as radiation) dominates
over any other form
of energy. During that period phase transitions described by particle physics
took place to give rise to hadron formation, baryogenesis, nucleosynthesis and so on.
Because in an expanding Universe radiation energy dilutes faster than
the energy of pressureless matter, in the second
 phase the latter becomes
dominant. Large scale structures (LSS) like galaxies and galaxy clusters
formed during that period. In both phases the Universe expands in a
decelerated fashion
(for a short review on the SBB see the contribution to this book by
J.\ L.\ Cervantes--Cota\cite{Cervantes} and for more details
the book \cite{inflation}).

The SBB can easily accommodate phases of accelerated expansion of the
Universe. According to cosmological observations, such a phase could
correspond to the present state of the observable Universe and seems to be
necessary in the very early Universe in order to solve several problems
inherent to the SBB, particularly
those problems related to the initial conditions.

For the Universe to expand acceleratingly, a very special kind of energy density
is required to dominate over the remaining contributions to the total energy
budget. This kind of energy is related to a
negative pressure. One of the outstanding problems in modern cosmology is
to find out what exactly this kind of matter is. It could be the case that
there are different explanations to
what causes the Universe to undergo cosmic acceleration in the present and in
the very early phases of its evolution. As to the present era, a dominating
vacuum energy is good enough to explain the observations but it
introduces other problems which seem to be very difficult to solve. A
good candidate is, instead,
a single scalar field with dynamics dominated by its
potential energy. This is also the favorite candidate to explain
an era of cosmic acceleration in the very early Universe. In both
cases, the problem is then to determine what the high-energy physics
framework is where such scalar fields arise.

In the next section the condition for an accelerated
FRW Universe will be derived. Then, in Sec.~\ref{sec:scalfield}
a scalar field will be described from the cosmological point of view,
along with how it could be used to induce an accelerated expansion.
Section \ref{sec:Obs2}
is devoted to explaining some ideas aimed at finding observational
signatures in the data, allowing us to distinguish the origin of cosmic
acceleration. If scalar fields are responsible for the observed and
predicted eras of accelerated expansion, the
observations should say something about high-energy physics that is
outside the scope of Earth-based laboratories. Finally, conclusions
are presented in section \ref{conclusions}.

Throughout this contribution natural units are used, i.e., $c=\hbar=1$.

\section{Accelerated Friedmann--Robertson--Walker Universe}
\label{sec:FRW}

Models in Physics are based on a set of principles derived from
observations and assumptions that make computations simpler
without wiping out crucial features of the phenomenon to be
modeled. The SBB is not an exception. The first assumption is that
there exist a \emph{cosmic scale}, quite a bit larger than the
galaxy clusters scale, $100-200$ Mpc. After averaging on those
scales, everything we observe in the sky dilutes into an isotropic
picture. This means that the Universe is the same when seen by a
terrestrial observer in any direction. But, according with Science
history, the human being does not seems to be such a special
being. Thus, it is assumed that we live in an ordinary planet
orbiting an ordinary star in a ordinary galaxy which is an
ordinary member of an ordinary galaxy cluster. It implies that any
observer located at any other point in the observable Universe
will see exactly the same picture of the sky, averaged in cosmic
scales, that a human observer does. In other words, the Universe
is assumed to be isotropic and homogeneous. This is known as the
\emph{cosmological principle}.

The physical distance $X_{phys}$ between two observers will change
with time, even if they are in relative rest each with respect to
the other,
\begin{equation}
\label{eq:distances}
\vec{X_{phys}}=a(t)\times\vec{X_{com}}\, .
\end{equation}
The physical distance between observers will be equal to the
distance between them if space does not expand (or contract),
i.e., $X_{com}$, times factor $a(t)$ which describes the change in
size of the expanding (contracting) homogeneous Universe after
given amount of time. The coordinate system where $X_{com}$ is
defined is called the comoving frame. Since the age of the
Universe is one of the quantities that can be inferred from the
observations, the homogeneity of the Universe must be defined on a
surface of constant proper time since the Big Bang. Time dilation
causes the proper time measured by an observer to depend on the
velocity of the observer, hence the time variable $t$ is actually
the proper time for comoving observers since the beginning of the
cosmological evolution. Distances in such a space--time are given
by the FRW metric; see Eq.~(3) in \cite{Cervantes}. There, the FRW
equations are deduced (see Eqs.~(4) and (5) in \cite{Cervantes}),
and solutions are found for a flat Universe (Eqs.~(7), (8), (9)
and (10) in \cite{Cervantes}). The Friedmann equation (Eq.~(4) in
\cite{Cervantes}) that determines the Hubble parameter can be
rewritten as,
\begin{equation}
\label{eq:FOE}
\Omega - 1 = \frac{k}{\dot{a}^2} = \frac{k}{a^2H^2}
 = k d_{Hcom}^2\, ,
\end{equation}
where a new definition is introduced, namely the \emph{comoving
Hubble radius} $d_{Hcom}\equiv d_H/a$, with $d_H \equiv H^{-1}$
being the \emph{physical Hubble radius} \footnote{Note that this
definition does not coincide with the one for the causal horizon
given in Refs.~\cite{Cervantes,Copeland} where $d_H\equiv
a(t)\int^t_{t_*}dt/a(t)$; They are different during an
inflationary era but are proportional to each other when the
expansion is of power-law type.}. Hence, $\Omega(\equiv
\rho/rho_c)$ can take values less, equal or greater than unity in
open, flat and closed Universes, respectively. From this equation
important conclusions can already be drawn about the differences
between accelerated and decelerated cosmologies. In the case of
cosmological evolution with $\ddot{a}>0$ ($\ddot{a}<0$) the
comoving Hubble radius decreases (increases) with time and
$\Omega$ converges to (diverges from) unity, implying that, with
time, the corresponding spatial hypersurfaces look more and more
(less and less) flat. The general condition for a universe to
expand or to contract acceleratingly can be drawn from Eq.~(5) in
\cite{Cervantes},
\begin{equation}
\label{eq:AC}
p<-\frac{\rho}{3}\, ,
\end{equation}
i.e., it must be filled with a fluid having a
sufficiently large negative pressure.

From the continuity equation (Eq.~(6) in \cite{Cervantes}) is not
difficult to see that if the energy density is constant, then
either the Universe is static (i.e., the scale factor $a$ does not
change in time) or the fluid satisfies the equation of state
$p=-\rho$. Since a particular value of this constant energy
density can be $\rho =0$, this energy is commonly associated with
the (scalable) energy of the cosmic vacuum. According to condition
(\ref{eq:AC}), the case of the non-static Universe with equation
of state $p_\Lambda=-\rho_\Lambda=-(8\pi G)^{-1}\Lambda={\rm
constant}$, implies that, whenever the weak condition
$\rho_\Lambda>0$ is satisfied, the Universe is described by an
accelerated expansion (or contraction) of the spatial
hypersurfaces.

Vacuum is a particular case of barotropic fluids with equation of
state, $p=w\rho $, where $w$ is, in general, a function of time.
Condition (\ref{eq:AC}) now reads, $w<-\frac{1}{3}$. For the case
of $w=constant$, the continuity equation yields, $\rho = \rho_0
a^{-3\left(1+w\right)}$.

The SBB includes two evolutionary stages. One extents from the Big
Bang until nearly the beginning of the epoch of galaxies
formation. To match several observations (the more important being
the abundances of light elements as predicted by nucleosynthesis),
the period had to be dominated by relativistic matter known, in
general, as radiation with $w_R=1/3$. After that period, the
formation of large--scale cosmological structure requires
 non--relativistic pressureless matter ($w_M=0$) to dominate over radiation.
Therefore, according to $w<-\frac{1}{3}$, the SBB describes a
Universe that expands non-acceleratingly from the very beginning
through the far future.

The SBB provides us a picture of an expanding Universe that
evolves from an initial singularity until today, passing through
the above--described $w$-epochs. It is successful in explaining
the formation of light elements (nucleosynthesis) and provides a
general framework to understand the evolution of perturbations
that eventually gave rise to the formation of LSS. However, as it
is pointed out in the contributions by J.\ L.\ Cervantes--Cota
\cite{Cervantes} and E.\ Copeland \cite{Copeland} to this book,
the SBB has unavoidable problems (horizon, flatness, causal origin
of primordial perturbations, etc) that cannot be understood
without the incorporation of new concepts and ideas. The main
ingredient that particle physics has brought to modern
cosmological understanding is that of scalar field dynamics. The
scalar field represents a generic matter field that evolves with
the Universe expansion and should be responsible for an
inflationary epoch at the very beginning of time, and perhaps
should also be responsible for the present accelerating dynamics
of the Universe. In the next section we study the dynamics of this
generic field.

\section{Scalar Fields}
\label{sec:scalfield}

A real scalar field is a map $\phi:\mathrm{M}\rightarrow
\mathrm{l\!R}$, i.e., a real function that puts a point in the
space--time ${\mathrm M}$ into relation with a point in the line
$\mathrm{l\!R}$. In Quantum Field Theory this function is used to
represent a boson particle. If the boson lives in a Minkowski
space--time ${\mathcal M^4}$, then the corresponding action is
given by (the details of the calculations in this section can be
found in \cite{inflation}),
\begin{equation}
\label{eq:SMphi}
S=\int_{\mathcal M^4} dx^4 {\mathcal L} \, ,
\end{equation}
with Lagrangian density,
\begin{equation}
\label{eq:LMphi} {\mathcal L} = -\frac 1 2
\eta^{\mu\nu}\phi_{,\mu}\phi_{,\nu} - V(\phi) \, ,
\end{equation}
where $\eta^{\mu\nu}=diag\{-1,1,1,1\}$ stands for the metric of
${\mathcal M^4}$, $\mu,\nu=0,1,2,3$ and $V(\phi)$ is the scalar
field potential. Varying the action, the equation of motion for
the scalar field is obtained as,
\begin{equation}
\label{eq:kgM} \ddot{\phi}-\nabla^2\phi + V^\prime(\phi)=0 \, ,
\end{equation}
where a prime denotes derivative with respect to $\phi$.

In the cosmological framework, ${\mathcal M^4}$ is substituted by
the ${\mathcal{FRW}}$ space--time and the action is that of
Einstein--Hilbert,
\begin{equation}
\label{eq:SRWphi}
S=\int_{\mathcal{RW}}dx^4 \sqrt{-g} {\mathcal L} \, ,
\end{equation}
where $g$ is the determinant of the FRW metric,
\begin{equation}
\label{eq:LRWphi}
{\mathcal L} = \frac 1 2 m^2_{\mathrm{Pl}} R
-\frac 1 2 g^{\mu\nu}\phi_{,\mu}\phi_{,\nu} - V^\prime(\phi)
\, ,
\end{equation}
$m_{\mathrm{Pl}}$ is the Planck mass and $R$ is the scalar curvature.
Equation (\ref{eq:kgM}) becomes,
\begin{equation}
\label{eq:kgRW}
\ddot{\phi} + 3H\dot{\phi} + V^\prime(\phi)=0 \, ,
\end{equation}
where a friction-like term arises due to the cosmic expansion and
the gradient terms were omitted, consistent with the cosmological
principle.

After comparison with the continuity equation (Eq.~(6) in
\cite{Cervantes}), the scalar field becomes equivalent to a
perfect fluid with energy density and pressure,
\begin{equation}
\label{eq:PRhophi}
\rho = \frac{\dot{\phi}^2}2 + V(\phi) \quad , \quad
p = \frac{\dot{\phi}^2}2 - V(\phi) \, .
\end{equation}
Hence, for a Universe dominated by the energy density of a real
scalar field, the condition (\ref{eq:AC}) for accelerated
expansion is rewritten as,
\begin{equation}
\label{eq:ACphi}
\frac{\dot{\phi}^2}2 < V(\phi) \, .
\end{equation}

With the aim of facilitating the analysis of cosmological
dynamics, it is convenient to define the horizon--flow functions
that are given in terms of the derivatives with respect to the
e--foldings $N\equiv \ln (a/a_i)$ of the Hubble horizon; the
latter is a basic ingredient to understand the causal evolution of
the cosmological dynamics; see the horizon problem in
\cite{Cervantes,Copeland}. Thus, the horizon--flow functions are
\cite{HFFampl}:
\begin{equation}
\label{eq:Hjf} \epsilon_0\equiv \frac{d_H(N)}{d_{Hi}}\, , \quad
\epsilon_{m+1}\equiv \frac{d\ln|\epsilon_m|}{dN}\, , \quad m\geq
0\, ,
\end{equation}
where $d_{Hi}$ is the value of the Hubble horizon at an arbitrary
initial $N_i$. Note that $\epsilon_1=\dot{d_H}$ and
$\epsilon_1\epsilon_2=d_H\ddot{d_H}$. According to condition
$\ddot{a}>0\rightarrow d\ln{d_H}/dN < 1$, for a positive energy
density, $0<\epsilon_1<1$ during an accelerated expansion.

Further, in accordance with definitions (\ref{eq:Hjf}),
\begin{equation}
\label{eq:H^2}
H^2(N) = H_0^2\exp\left(-2\int \epsilon_1(N) dN \right)
\, ,
\end{equation}
where $H_0$ is an integration constant. Now, substituting $\rho$
as given by (\ref{eq:PRhophi}) in the Friedmann equation (Eq.~(4)
in \cite{Cervantes}) and using the definition (\ref{eq:Hjf}) for
$\epsilon_1$ \cite{inflation,Terrero-Escalante:2001rt},
\begin{equation}
\label{eq:dPhidN}
\frac{d\phi}{dN}=\sqrt{\frac 2 \kappa}\sqrt{\epsilon_1} \, ,
\end{equation}
where
$\kappa\equiv 8\pi G=8\pi/m^2_{\rm Pl}$.
Given a scalar field cosmology characterized by $\epsilon_1(N)$ the
corresponding potential as function of the field is given by,
\begin{equation}
\label{eq:Vphi}
V(\phi)= \left\{
\begin{array}{rcl}
\phi(N)&=& \sqrt{\frac2\kappa}{\displaystyle \int}
\sqrt{\epsilon_1} dN - \phi_0
\, ,
\\
V(N)&=&\frac{H_0^2}\kappa\left[3-\epsilon_1\right]
\exp\left(-2{\displaystyle \int} \epsilon_1 dN \right)
\, ,
\end{array}
\right.
\label{eq:C2Vp}
\end{equation}
where the potential as function of $N$
is derived from the Friedmann
equation for the scalar field cosmology using expressions
(\ref{eq:dPhidN}) and (\ref{eq:H^2})
\cite{inflation,Ayon-Beato:2000}.

\section{Observations and Modeling}
\label{sec:Obs2}

\subsection{Present Day Acceleration}
\label{ssec:quintessence}

Recent observations of the celestial candles known as type Ia
supernovae have been made that indicate a new feature of the
present Universe composition. Currently, the physics behind the
peak light output from such supernovae seems to be well
understood. Thus, by observing a type Ia supernova in a distant
galaxy, measuring the peak light output, an comparing the relative
intensity of light observed from the object with that expected
from its absolute magnitude, the inverse square law for light
intensity can be used to infer its distance. Because type Ia
supernovae are very bright objects, they are used to measure
distances out to around 1000 Mpc, which is a significant fraction
of the radius of the observable Universe. According to the
analysis of the data collected for several type Ia supernovae, the
observable Universe seems to be in a phase of accelerated
expansion; for details on the type Ia supernovae and on the
analysis of the redshift data see the contribution by A.\
Filippenko to this book \cite{Filippenko}.

Therefore, in accordance with Eq.~(\ref{eq:AC}), the Universe
would be currently dominated not by pressureless matter but by
some kind of fluid with negative pressure. Since this component of
the cosmic energetic budget has eluded direct observation so far,
it is generically known as \emph{dark energy} (see the
contributions to this book by Copeland \cite{Copeland} and de la
Macorra \cite{Macorra}). The dark energy can be, in principle, the
non-zero vacuum energy parametrized by the cosmological constant
$\Lambda$. Adding such an energy does not strongly modify the
cosmological picture as described by the SBB. In fact, the dark
energy seems also to be necessary in order to match data from the
observation of cosmic microwave background and from large--scale
structure formation. This fact is a very interesting confirmation
of the existence of the dark energy.

First of all, if one would like to describe the present time
cosmic acceleration as induced by a cosmological constant, the
associated vacuum energy required to match the observations is
$\rho^{Obs}_{\Lambda}\le (10^{-12} GeV)^4$. On the other hand,
from Quantum Field Theory with cutoff at the Planck scale it is
expected that $\rho^{QFT}_{\Lambda}\sim (10^{18} GeV)^4$. The
large disagreement between the two estimations of the vacuum
energy is one of the hottest problem of modern cosmology. This is
one of the aspect of the so--called \emph{cosmological constant
problem}.

A solution to this problem is to consider a cosmological
``constant'' decaying in such a way that its associated energy is
currently the one required by observations. According to
Eq.~(\ref{eq:PRhophi}), the simplest candidate for such a
dynamical vacuum energy could be a scalar field slowly changing
over time. For a high enough potential energy, condition
(\ref{eq:ACphi}) is fulfilled and the Universe permeated by the
scalar field potential energy undergoes accelerated expansion.
This scalar field has been coined \emph{quintessence}.

\begin{figure}
\centering
\includegraphics[height=6cm]{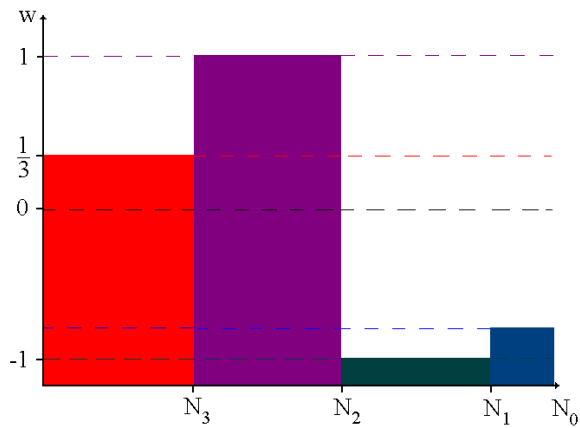}
\caption{Parameterization of the dark energy evolution}
\label{fig:wparam}
\end{figure}

Many quintessence models have been devised and
some other candidates for the dark energy have been proposed like the
\emph{Cardassian expansion} \cite{Freese:2002sq}
and the \emph{Chaplygin gas} \cite{Kamenshchik:2001cp}.
Among the
problems
that arise when modeling the dark energy, one of the outstanding ones is to
find observational signatures differentiating between the candidates.
One expects the observations to help in that task but often it is necessary
to know exactly what to look for in data. Since different dark energy
candidates evolves in different ways, it could be useful to look for
the imprint of these differences in data.
With that aim, a model independent
parameterization of the dark energy evolution can be useful.
In figure \ref{fig:wparam}
it is shown how this parameterization can be done. In the
horizontal axis the e--folds number are calculated now with
respect to the present time $N_0$, i.e., $N$ increases leftward.
In the vertical axis is $w(N)$, which determines the equation of
state $p=w\rho$. With dashed lines are denoted four cases where
$w$ is a constant: a stiff fluid ($w_s=1$), radiation ($w_R=1/3$),
matter ($w_M=0$), a tracker scalar field (which is a special case
of quintessence models where $w$ can be safely approximated to be
constant with $-1<w_{tracker}<-1/3$), and a cosmological constant
($w_\Lambda=-1$). Those cases that do not satisfy the condition
$w<-1/3$ get thrown out of the game because they do not produce an
accelerated expansion.

More generally, one can look for parameterizations where $w$ can
be approximated by different constants for different ranges of
e--folds number. In the figure the filled polygons represents
case,
\begin{equation}
\label{eq:wpar}
w(N) =
\left\{ \begin{array}{l}
\quad \quad \;\;w_4=\frac 1 3, \quad \quad \quad \quad \,\, N > N_3 \\
\quad \quad \;\;w_3=1, \quad \quad N_3\ge N > N_2 \\
\quad \quad \;\;w_2=-1, \quad  \, N_2\ge N > N_1 \\
               -1<w_1<-\frac 1 3, \quad N_1\ge N > N_0
\end{array} \right.
\end{equation}
where $-1<w_1<-1/3$.  As it is explained in
\cite{Macorra},
this could be the case of some fluid which initially
behaves like radiation, at some energy scale condensates into a scalar field
with dynamics dominated by its kinetic energy, then undergoes a strong friction
changing to a phase of totally potential dominated dynamics and, finally,
behaves like a tracker field.

It is interesting to check, for instance, how such a
parameterization will modify the CMB spectrum and how it compares
to a cosmological constant and to tracker fields. CAMB is a
program which permits one to compute the CMB spectrum after
specifying a number of parameters including $w_{tracker}$
\cite{CAMB}. To include the constant in sectors parameterization
of $w(N)$ it was necessary to match the values of the energy
densities $\rho_{wi}$ at the borders of those sectors,
\begin{equation}
\label{eq:RHOpar}
\rho_{wi}(N)=\rho^{eff}_{i}\exp\left[-3(1+w_i)N\right]\, ,
\end{equation}
where,
\begin{equation}
\label{eq:RHOeff}
\rho^{eff}_{i}=\rho_0 \exp\left(-3\Delta w_{12} N_2\right)
\exp\left(-3\Delta w_{23}N_3\right)\dots
\exp\left(-3\Delta w_{(i-1)i}N_i\right)
\end{equation}
and $\Delta w_{(i-1)i}=w_{(i-1)}-w_i$.

\begin{figure}
\centering
\includegraphics[height=12cm,angle=-90]{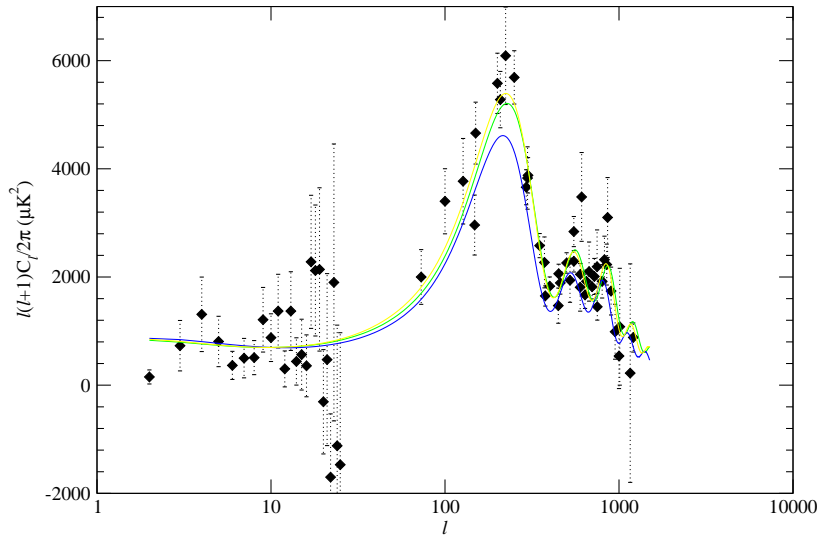}
\caption{The CMB spectra for different versions of the
parametrization by sectors of the state parameter $w$.}
\label{fig:exs}
\end{figure}

\begin{figure}
\centering
\includegraphics[height=12cm,angle=-90]{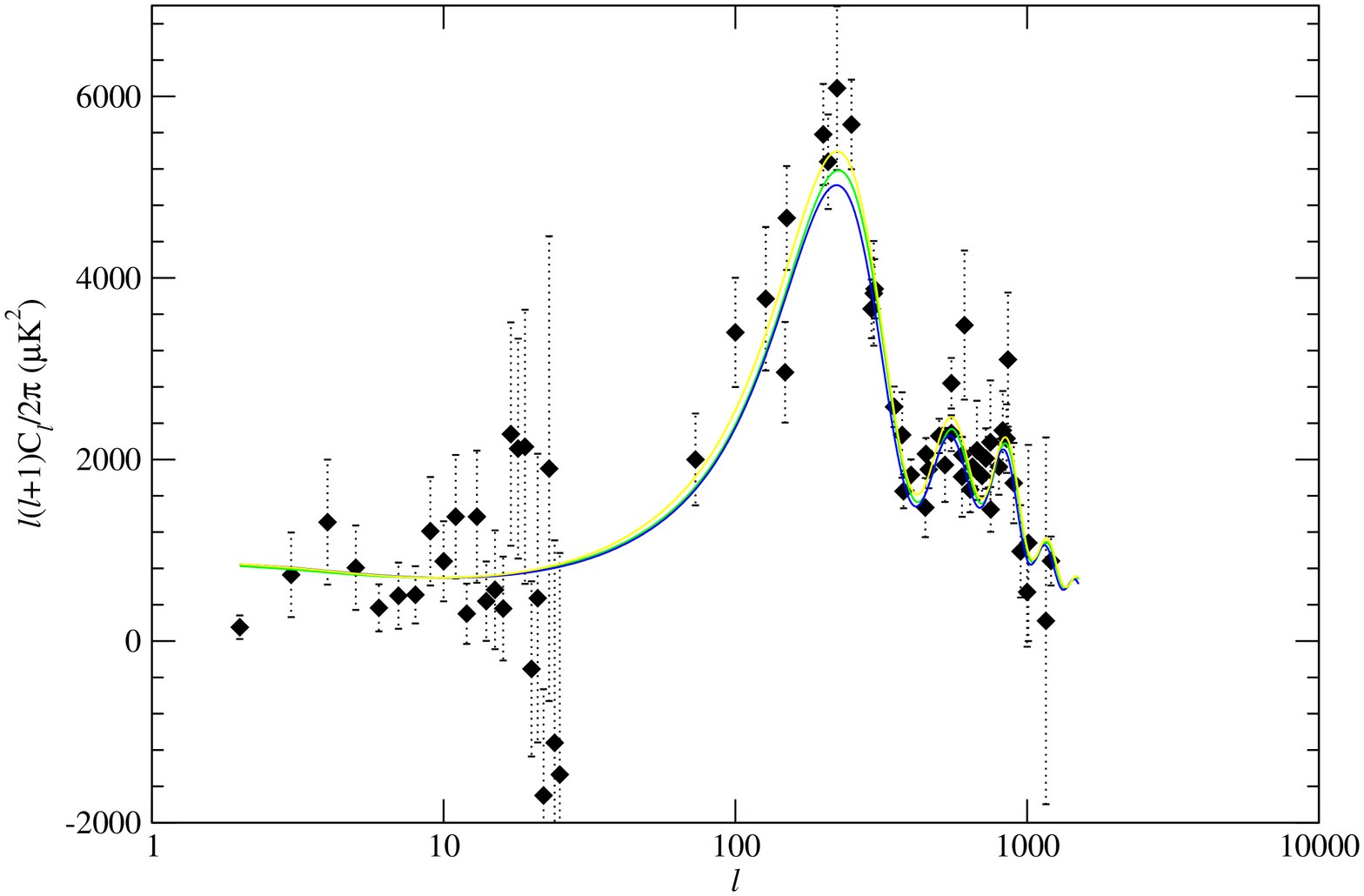}
\caption{Comparing a tracker field case (blue curve),
the $\Lambda$ case (green curve) and a case with $w(N)$
parametrized by sectors (yellow line).}
\label{fig:sectors}
\end{figure}

In figure
\ref{fig:exs},
examples are presented of the variations of the CMB spectrum after
varying the parameters in (\ref{eq:wpar}). In this figure the
binned data from experiments DASI, Boomerang, MAXIMA and COBE--DMR
is also shown. Though a more detailed analysis is still in
process, it can be already noted that variation of the values of
$N_1$, $N_2$ and $N_3$ modifies the heights and positions of the
peaks and dips of the theoretical curve of the CMB spectrum. In
fact, one can choose the values of these parameters in such a way
that the fit to data might be improved compared to the cases of a
cosmological constant and a tracker field as shown in figure
\ref{fig:sectors},
where the blue curve represents the spectrum for a tracker field with
$w_{tracker}=-0.86$, the green curve stands for a cosmological constant and
the yellow one corresponds to $N_1=1.97$, $N_2=4.2$, $N_3=6.5$ and
$w_1=-0.86$. Certainly, the data error bars are still too large to make any
strong conclusion in this direction but results are encouraging.

\subsection{Cosmic Acceleration in the Very Early Universe}
\label{ssec:veu}

As in the case of the present-day Universe, if the acceleration in
the very early times --required to solve the SBB problems-- is
proposed to be induced by the vacuum energy associated with a
cosmological constant, then several problems are confronted. The
theory of nucleosynthesis, which is so successful in accounting
for the observed cosmic abundances of the lightest atoms,
describes a radiation--dominated Universe. In this way, in order
to reproduce the SBB success, the inflationary period must end
some time before nucleosynthesis started. Moreover, if the density
perturbations leading to large--scale structure are wanted to be
causally seeded by an inflationary period in the very early
Universe, then the Hubble radius at some point must start to
increase, which implies the end of the accelerated expansion.

Once the energy of a cosmological constant begins to dominate over
other non-exotic forms of energy, it will dominate forever. Thus,
the inflationary epoch will have no end. Once more, a single
scalar field seems to be the simplest candidate to solve all the
problems associated with a cosmological constant while still
producing enough inflation. In this framework, the origin of the
density perturbations leading to LSS is thought to be the quantum
fluctuations of the scalar field during inflation
\cite{inflation,Bran}. A distinguishing feature of this scenario
is that, along with the density perturbations associated with the
inflaton field, tensor perturbations are also produced which are
associated with space--time metric perturbations. All of these
fluctuations will be amplified by the accelerated expansion and
seeded at the required time through the mechanism of crossing out
and crossing back the Hubble horizon. The primordial perturbations
are parameterized as,
\begin{eqnarray}
\label{eq:SExp}
\ln A^2(k)&=&\ln A^2(k_*) + n(k_*)\ln\frac{k}{k_*}
\nonumber\\
&+& \frac12\frac{d\,n(k)}{d\ln k}\vert_{k=k_*}\ln^2\frac{k}{k_*}+\cdots
\, ,
\end{eqnarray}
where $A$ stands for the normalized amplitudes of the scalar ($A_S$)
or tensor ($A_T$)
perturbations, the corresponding spectral indices, $n$,
are defined by,
\begin{eqnarray}
\label{eq:nSDef}
n_S-1&\equiv&\frac{d\ln A_S^2}{d\ln k} \, ,\\
\label{eq:nTDef}
n_T&\equiv&\frac{d\ln A_T^2}{d\ln k} \, ,
\end{eqnarray}
and $k=aH$ is the comoving wavenumber corresponding to the
wavelength matching the Hubble distance. As it was discussed in
the previous chapters of this book, one of the problems of the SBB
is the very constrained nature of the initial conditions required
for the LSS formation. One of the most restricting requirement is
the almost \emph{scale invariant} nature of the primordial
perturbations. Obviously, this impose strong constraints on the
values of the parameters in expansion (\ref{eq:SExp}). Along with
the requirement of yielding a sufficiently large number of
e--foldings, $n_S\approx 1$ and $n_T\approx 0$ are strong
conditions to determine whether a proposed scalar field model is a
successful inflationary model.

The horizon--flow functions (\ref{eq:Hjf})
serve as link between observations and the
inflationary dynamics.
For most inflationary models, exact expressions for parameters in
(\ref{eq:SExp}) are unknown. The usual way to calculate them is as an
expansion in terms of the horizon--flow functions
(see \cite{Stewart:1993bc,HFFampl} for details).
To next--to--leading order
in
terms of the horizon--flow functions, indices
(\ref{eq:nSDef}) and (\ref{eq:nTDef})
are written as
\cite{Stewart:1993bc,Terrero-Escalante:2002sd}
\begin{eqnarray}
\label{eq:SLns}
n_S-1&=& -2\epsilon_1 - \epsilon_2 - 2\epsilon_1^2 - (2C+3)\epsilon_1\epsilon_2
- C\epsilon_2\epsilon_3, \\
\label{eq:SLnt}
n_T&=& -2\epsilon_1 - 2\epsilon_1^2 - 2(C+1)\epsilon_1\epsilon_2 \, ,
\end{eqnarray}
where $C\approx-0.7293$. Dynamics described using the
horizon--flow functions correspond to an inflationary potential
correspond. This way, the corresponding spectral indices can be
calculated and compared with the observational values.

The big problem here is that there exists a large number of single
scalar fields models in good agreement with observations,
therefore it is difficult to determine which is the one
corresponding to the actual physics in the very early Universe.
With this aim, a more efficient approach seems to be to constrain
and determine the main features of the inflaton potential
according to observations, instead of building models and
comparing their predictions with the measured data (see
\cite{Copeland} and \cite{Easther:2002rw} for references on this
approach).

Recalling that, according with (\ref{eq:Hjf}),
the definition of $\epsilon_{m+1}$ involves the derivative with respect to
$N$ of $\epsilon_m$,  (\ref{eq:SLns}) and (\ref{eq:SLnt}) are therefore
differential equations for $\epsilon_1$.
In this way, solving these equations and
using expression (\ref{eq:Vphi}), the inflaton potential
can be determined from the information
on the functional forms of the tensor and scalar
spectral indices.

The strong limitation for this program to be useful is that the
most that is known (and will be for a while) about the scale or
time dependence of the spectral indices is the observed values of
a very few parameters in expansions (\ref{eq:SExp}), together with
the corresponding error bars. Taking this limitation into account,
the best one can do is to look for generic features of the
potential yielding values of the primordial parameters in
agreement with those derived from observations, using some ``well
based'' assumptions for the values of those parameters which have
not been observed so far.

For instance, in \cite{ConstNt} it was proved that if it is assumed
$dn/d\ln k=0$ for the scalar and the tensor perturbations, then the resulting
potential is an exponential function of the inflaton field. This corresponds
to the scenario known as \emph{power-law inflation} because $a\sim t^p$ with
$p\gg 1$ \cite{PLinfl}.

If $dn_S/d\ln k\ll 1$ is allowed to be non-zero while keeping
$dn_T/d\ln k=0$, it can be seen that power--law inflation is an
attractor of the corresponding inflationary dynamics
\cite{ConstNt}. This implies that it is difficult to distinguish
the actual potential from the exponential one using only the
observational information.

A similar result is obtained if both spectral indices are allowed
to be scale--dependent but with this dependence being detectible
up to the second order on the horizon--flow functions
\cite{Terrero-Escalante:2001rt}.

The role of the tensor perturbations deserves special attention
when determining the best--fit values of the cosmological
parameters from CMB and LSS spectra. That is motivated in part by
the possibility of measuring the cosmic background polarization
\cite{inflation}, allowing the tensorial contribution to be
indirectly determined. This contribution can be parametrized in
terms of the relative amplitudes of the tensor and scalar
perturbations,
\begin{equation}
\label{eq:r}
r \equiv \alpha \frac{A_T^2}{A_S^2}\, ,
\end{equation}
where $\alpha$ is a constant. The expectation is to measure a
central value of $r$. Thus the question is how look like the
inflationary dynamics yielding an almost constant ratio $r$. The
answer was given in \cite{Terrero-Escalante:2001du} where it was
shown that in the case of an exactly constant $r$, power--law
inflation is a repulsor of the corresponding dynamics. Since it
describes a quick and strong depart from scale--invariance, for
the model to be successful, the perturbations must be produced in
the quasi power--law regime, once more making it difficult to
observationally distinguish between the two dynamics.

All of the above-discussed results imply a serious handicap for
any program of reconstruction of the inflaton potential. A way to
improve this situation, could be to combine the information on
$\Delta n$ (the difference between the spectral indices) and the
value of $r$, with the two first horizon--flow functions
\cite{Terrero-Escalante:2002uj}. It follows from definitions
(\ref{eq:nSDef}), (\ref{eq:nTDef}) and (\ref{eq:r}) that
\begin{equation}
\label{eq:dlnr/dlnk}
\frac{d\ln r}{d\ln k}= \Delta n \equiv n_T - (n_S-1) \, ,
\end{equation}
this way, any information on the evolution of both spectral indices can be
used as information on the scale dependence of the tensor to scalar ratio.

\begin{figure}
\centering
\includegraphics[height=6cm]{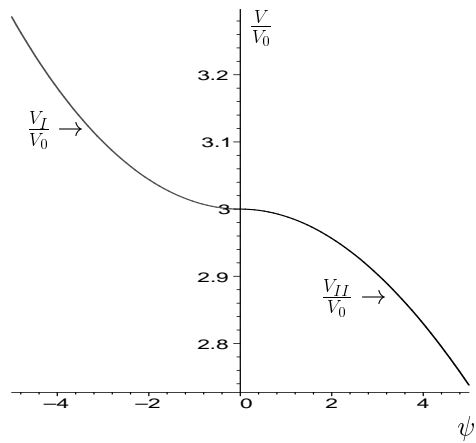}
\caption{Sectors of the inflaton potential given by (\ref{eq:Vai}) for
$A=-0.0073$ ($V_I$) and $A=0.0073$ ($V_{II}$).}
\label{fig:rDynV}
\end{figure}

With regards to these, it becomes important to analyze in details
the case,
\begin{equation}
\label{eq:ra1}
\ln\frac{r}{16} = a_0 + a_1 (N-N_0)\, ,
\end{equation}
where the corresponding solution for $\epsilon_1$ is
\cite{Terrero-Escalante:2002uj},
\begin{equation}
\label{eq:e1a1}
\epsilon_1(N) = \epsilon_{1(i)}
\exp\left[B\exp\left(-\frac N C\right)\right]\exp\left(A N\right)
\, ,
\end{equation}
with $\epsilon_{1(i)}\equiv \epsilon_{1(0)}
\exp\{\left[a_0+\left(N_0-C\right)a_1\right]C\}$ and $A\equiv a_1
C$. The asymptotes of this solution for $B\neq 0$ will be mainly
determined by the value and sign of $B$. However, for $A=0$ (i.e.,
$\ln r/16=a_0$), if the model yielding $\epsilon_1$ given by
(\ref{eq:e1a1}) is expected to be compatible with current data,
$B$ has to be chosen an extraordinarily small number. Once more,
this will make it very difficult to observationally distinguish
the corresponding scenario from power--law inflation. More
interesting is the case $B=0$ where the potential is,
\begin{equation}
\label{eq:Vai}
V = V_0\left(3-\frac {A^2} 4 \psi^2\right)\exp\left(-\frac A 2 \psi^2\right)\,,
\end{equation}
with $\psi\equiv \sqrt{\kappa/2}(\phi+\phi_0)$. In Fig.~\ref{fig:rDynV}
sectors of this potential are plotted for $A=-0.0073$ ($V_I$) and
$A=0.0073$ ($V_{II}$). For this model, information on the
existence of extrema and on the curvature can be derived. As can
be seen in Fig.~\ref{fig:rDynV}, realizations of this potential
resemble the cases of monomial potentials with even order ($V_I$,
$\epsilon_2<0$), and inflation near a maximum ($V_{II}$,
$\epsilon_2>0$) (see \cite{inflation} for examples of such
inflationary scenarios motivated by particles physics) allowing,
therefore, to observe features of the inflaton potential beyond
the exponential form characteristic of power--law inflation. In
\cite{Terrero-Escalante:2002uj} it was shown that for a large set
of $A$ and $\epsilon_{1(i)}$ values, the corresponding spectra
agree with CMB and LSS observations.

\subsubsection{On the Order of the Approximations}
\label{sssec:orders}

A crucial question in the analysis in the previous section is to
what extent these results depend on the order of expressions
underlying the calculations. It is widely believed that this is
not of concern. Let us show that it must be.

\begin{figure}
\centering
\includegraphics[height=5cm]{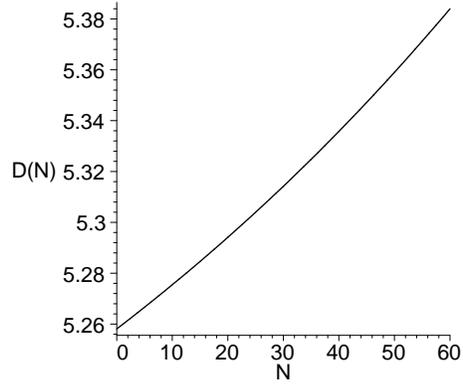}
\caption{The behavior of D(N) for $A=-0.0073$ and $\epsilon_{1(i)}=0.05$.}
\label{fig:damping}
\end{figure}

\begin{figure}
\centering
\includegraphics[height=5cm]{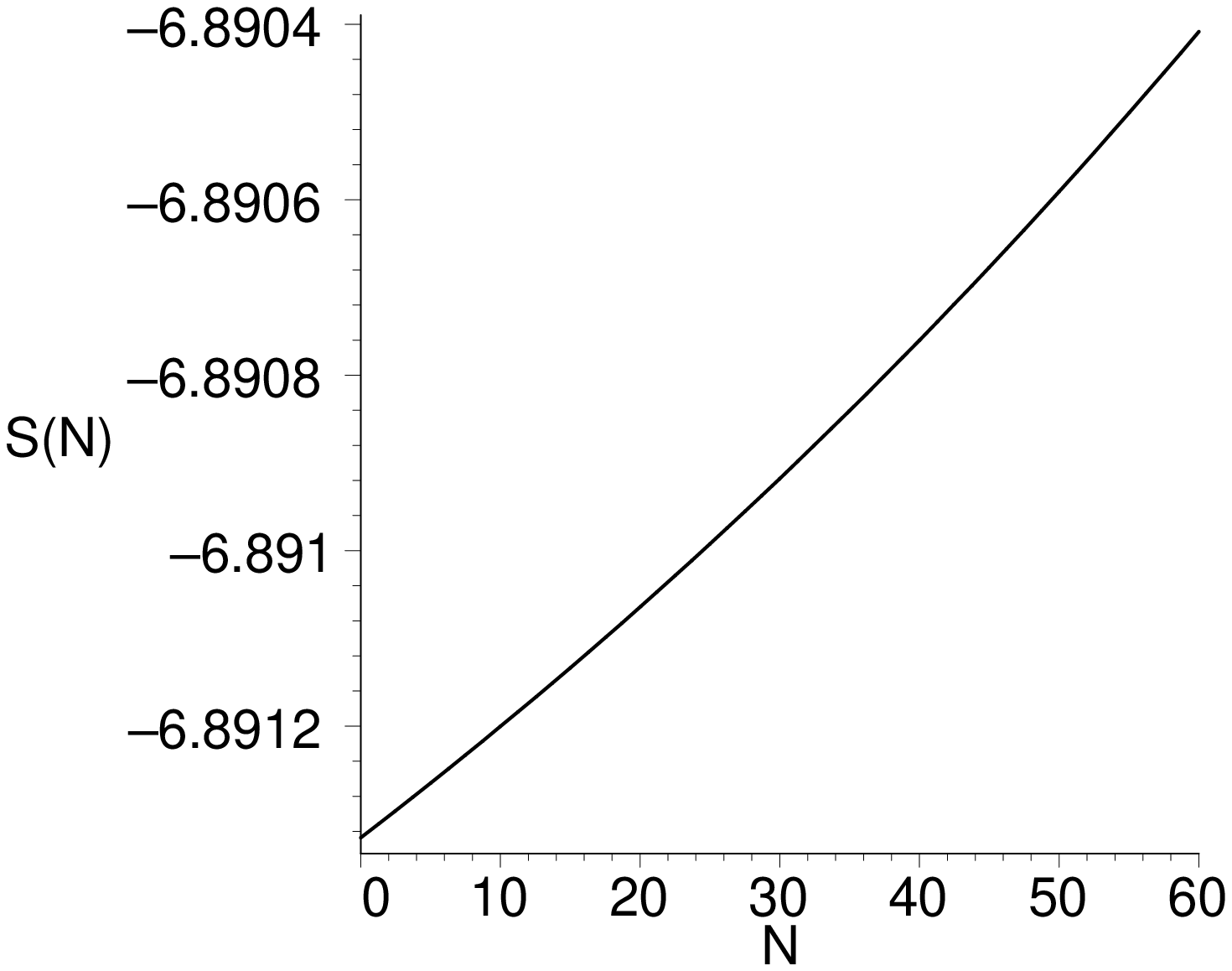}
\caption{The behavior of S(N) for $A=-0.0073$ and $\epsilon_{1(i)}=0.05$.}
\label{fig:stiffness}
\end{figure}

\begin{figure}
\centering
\includegraphics[height=5cm]{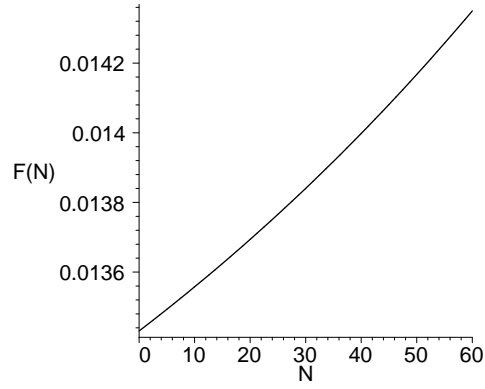}
\caption{The behavior of F(N) for $A=-0.0073$ and $\epsilon_{1(i)}=0.05$.}
\label{fig:forcing}
\end{figure}

To next--to--next--to--leading order the tensor to scalar ratio is
given by,
\begin{eqnarray}
\label{eq:NNLOr_r0}
\ln\frac{r}{r_0} &=& \ln\epsilon_1 + C\epsilon_2
+ \left(-\frac{\pi^2}2 + 5 + C \right)\epsilon_1\epsilon_2 \nonumber\\
&+& \left(-\frac{\pi^2}8 + 1\right)\epsilon_2^2 +
\left(-\frac{\pi^2}{24} + \frac{C^2}2\right)\epsilon_2\epsilon_3 \, .
\end{eqnarray}
If the order of this expression was of little concern, then the corresponding
solution for the case with $r$ given by (\ref{eq:ra1})
must be very similar to potential (\ref{eq:Vai}). To check this, one can
assume,
\begin{equation}
\label{eq:e1pert}
\epsilon_1 = \epsilon_1^{nlo}\left(1+\delta\right)\, ,
\end{equation}
with $\delta<<1$, yielding
\begin{eqnarray}
\label{eq:e2pert}
\epsilon_2 &=& \epsilon_2^{nlo} + \frac{d\delta}{dN}\\
\label{eq:e23pert}
\epsilon_2\epsilon_3 &=& \epsilon_2^{nlo}\epsilon_3^{nlo}
+ \frac{d^2\delta}{dN^2}
\, ,
\end{eqnarray}
where the super-index $nlo$ stands for the next-to-leading order
solution. $\delta(N)$ is expected to remain very small as $N$
increases. Substituting (\ref{eq:e1pert}), (\ref{eq:e2pert}) and
(\ref{eq:e23pert}) into (\ref{eq:NNLOr_r0}) with
$\epsilon_1^{nlo}$ given by expression (\ref{eq:e1a1}) it is
obtained that
\begin{eqnarray}
\label{eq:osc}
\frac{d^2\delta}{dN^2} &+& D(N)\frac{d\delta}{dN} + S(N)\delta = F(N)\\
D(N) &=& C_3+ C_2\epsilon_{1(i)}\,{e^{B{e^{N C_1}}}}{e^{AN}}\\
S(N) &=& -C_4+ C_2\epsilon_{1(i)}\,{e^{B{e^{N C_1}}}}{e^{AN}}
\left(  C_1 B{e^{N C_1}}+A \right)\\
F(N) &=&  C_2\epsilon_{1(i)}\,{e^{B{e^{N C_1}}}}{e^{AN}}
\left( C_1 B{e^{N C_1}}+A \right)+ C_5 B{e^{N C1}}+ C_6 A
\, ,
\end{eqnarray}
where non-linear terms of $\delta$ were neglected and $C_i$ (with
$i=1\dots 6$) are fixed constants given in terms of $C$ and $\pi$.
Constants $\epsilon_{1(i)}$, $A$ and $B$ are those already given
in the next--to--leading order solution (\ref{eq:e1a1}). In Figs.
\ref{fig:damping}, \ref{fig:stiffness} and \ref{fig:forcing} the
$N$-dependent parameters $D(N)$, $S(N)$ and $F(N)$ are plotted for
the same numerical values for the constants used in
\cite{Terrero-Escalante:2002uj}.

\begin{figure}
\centering
\includegraphics[width=6cm]{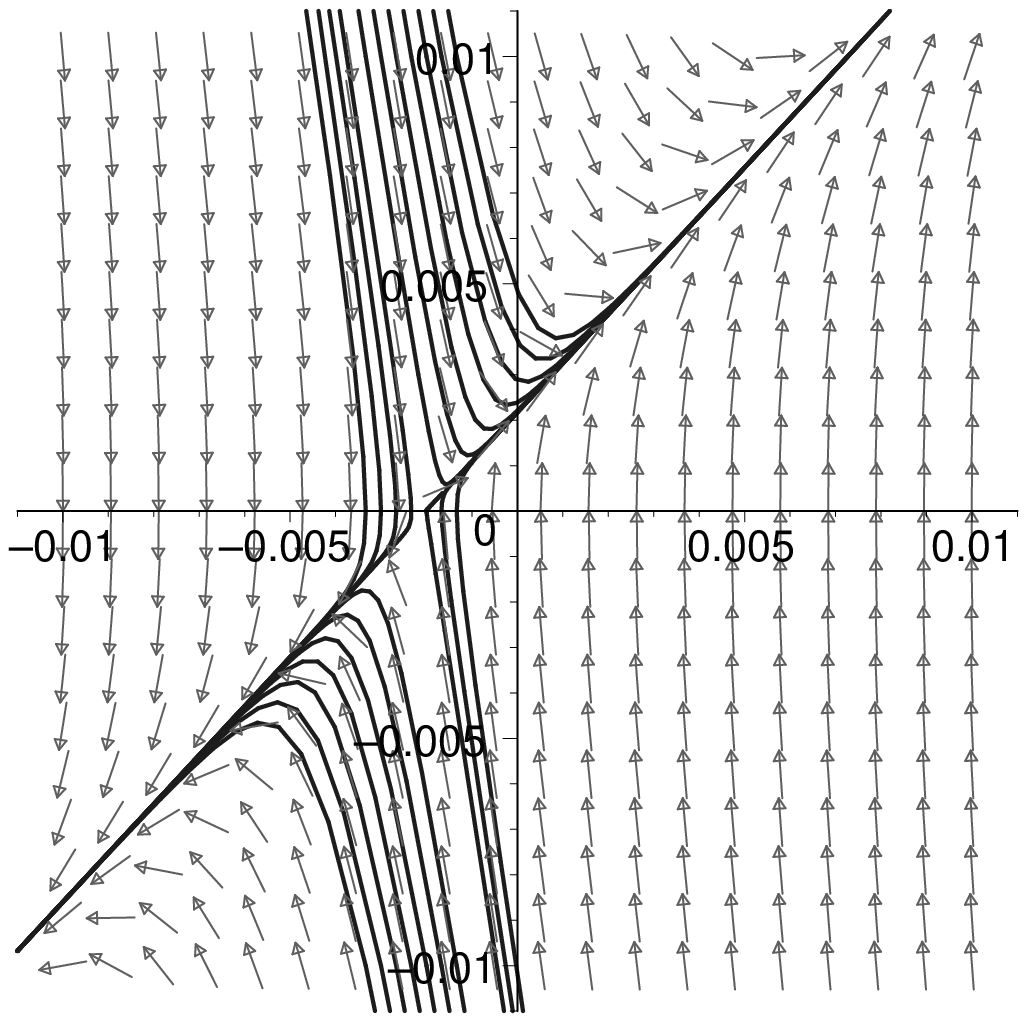}
\caption{The phase portrait for (\ref{eq:osc}) with
$D=-5.32$, $S=6.891$ and $F=0.014$. In the horizontal axis $\delta$ is
represented;
in the vertical axis, $\dot{\delta}$. Arrows show the vector field and
lines some trajectories.}
\label{fig:ppPert}
\end{figure}

At least for this case, the parameters can be safely approximated
by constants. Therefore, a qualitative analysis of the phase space
for the flow given by (\ref{eq:osc}) can be carried out. It yields
that there exists a saddle point at $(\delta=F/S,
\dot{\delta}=0)$, shown in figure  \ref{fig:ppPert}.
It means that solutions for $\delta(N)$ with $A=-0.0073$ and
$\epsilon_{1(i)}=0.05$ are likely unstable. A more complete
analysis is indeed required, but this result would serve as a
warning to pay more attention to the order of the approximations
used when information on the inflationary dynamics is drawn from
observational data.

\section{Conclusions}
\label{conclusions}

Cosmic acceleration is a trivial solution to the Einstein
equations for an isotropic and homogeneous Universe, as appears to
be the one where we live. An accelerated expansion is typical of
Universes filled with a kind of energy yielding a strong enough
negative pressure. This is the case when the cosmic energy is
dominated by the contribution of the vacuum.

Observational evidence strongly suggest that our Universe's
evolution includes three well-defined epochs with regards to the
increase of the cosmic volume. First, the very early Universe
would undergo an accelerated expansion known as inflation. Then, a
period of non-accelerated expansion would take place where most of
the known kinds of matter and matter structures were formed.
Finally, in recent times (with respect to cosmic scales) the
Universe would enter a second epoch of accelerated expansion where
the corresponding dominated matter-energy content is called dark
energy.

A real scalar field is a good candidate for inducing cosmic
acceleration. It may help to solve problems arising when
a constant vacuum energy is used to explain inflation or the nature of the
dark energy. For the required negative pressure, the
scalar field dynamics must be dominated by its potential energy.

A hot question in cosmology is whether the observed (predicted)
cosmic acceleration is (was) induced by a scalar field. If this is
the case, the relevant question is to determine the origin and
nature of the corresponding potential. This will open an important
window into high energy physics.

Here, an idea was hinted at on how to differentiate between
candidates for the dark energy. The proposal is to divide the
evolution of the dark energy in periods where the corresponding
equation of state could be approximated to be linear. The
best--fit values for the corresponding slopes would indicate the
favorite candidate. Encouraging results have been obtained in this
direction.

It was also explained some of the difficulties that arise when
deriving the inflationary potential from observations. It seems
like the best that can be done is to indicate generic features of
the potentials yielding perturbations spectra matching the
measured data. It was emphasized that the use of data on the
difference of the tensor and scalar indices of perturbations
yields information on the scale--dependence of the tensor to
scalar ratio of primordial perturbation amplitudes. This
information may be very useful in classifying the inflationary
potentials. Finally, a warning was issued about the possibility
that the features of the inflaton potential drawn from the
observational data could be biased by the order of the
approximations used to derive the expressions underlying the
calculations.

\subsubsection{Acknowledgments}
I would like to thank School organizers for inviting me to give
this talk. I also thank A.\ Garc\'{\i}a, A.\ de\ la\ Macorra and
A.\ Coley for their solid support and helpful discussions.
Supported in part by CONACyT grant 38495--E and SNI (Mexico).

%
%
%
%

%
%



\end{document}